%% file: metj_fulldata_v5.3.tex
\RequirePackage{lineno}
\documentclass[aps,prd,10pt,twocolumn,superscriptaddress,showpacs,nobibnotes,amsmath,amssymb]{revtex4-1}

\usepackage{comment}
\usepackage{multirow}
\usepackage{dcolumn}
\usepackage{bm}
\usepackage{lineno}  
\usepackage{setspace}
\usepackage{graphicx}
\includecomment{printrobustnotes}
\includecomment{printallnotes}
\usepackage{xspace}

\input{prdDefs.tex}

\begin{document}



\title{Top-quark mass measurement in events with jets and missing transverse energy using the full CDF data set}

\input{November2012_Authors.tex}


\date{\today}
\begin{abstract}
	We present a measurement of the top-quark mass using the full data set of Tevatron $\sqrt{s} = 1.96$~TeV proton-antiproton collisions recorded by the CDF II detector, corresponding to an integrated luminosity of \invfb{8.7}. 
The analysis uses events with one semileptonic $t$ or $\bar{t}$ decay, but without detection of the electron or muon. 
We select events with significant missing transverse energy and multiple jets. We veto events containing identified electrons or muons.
We obtain distributions of the top-quark masses and the invariant mass of the two jets from $W$-boson decays from data and compare these to templates derived from signal and background samples to extract the top-quark mass and the energy scale of the calorimeter jets with {\it in situ} calibration. 
A likelihood fit of the templates from signal and background events to the data yields the top-quark mass, $\mtop = \gevcc{\measStatSyst{173.93}{1.64}{0.87}}$. This result is the most precise measurement to date of the mass of the top quark in this event topology. 
\end{abstract}
\maketitle
The top quark~($t$) is the heaviest known elementary particle. Its mass is approximately 40 times larger than the mass of its isospin partner, the bottom quark~($b$). The top-quark mass~(\mtop) is a fundamental parameter of the standard model~(SM), and is tightly related to the $W$-boson mass and Higgs-boson mass via electroweak radiative corrections~\cite{hmass}.  Prior to the recent observation of the Higgs boson and a direct measurement of its mass~\cite{hobserv}, precision measurements of \mtop and $W$-boson mass provided the only available information on the SM Higgs boson.

Top quarks at the Tevatron are predominatly produced in \ttbar pairs. Assuming unitarity of the three-generation quark-mixing matrix~\cite{ckm}, the top quark decays almost exclusively into a $W$-boson and a $b$-quark.  The case where one $W$ decays leptonically into a charged lepton~($e, \mu, \tau$) and its neutrino and the other $W$ decays hadronically into a pair of jets ($\ttbar\rightarrow l\nu b\bar{b} q\bar{q}$) defines the lepton+jets decay mode. In the standard selection of lepton+jets events~\cite{ljcross,tevaverage}, we require a well-reconstructed electron or muon with multiple jets and large missing transverse energy~(\mets)~\cite{DefMET}.  The first requirement excludes events with a hadronically decaying $\tau$ lepton and events with an electron or muon that fails the identification requirements or falls outside the limited detector coverage.  In this paper, we focus on events from the lepton+jets decay in which no muon or electron are reconstructed. The signal acceptance in this channel is comparable with the standard lepton+jets channel, and the dominant QCD multijet background is manageable with a multivariate technique~\cite{METJcross, earlyMETJ}.  This work is an update of a previous measurement that used a subset of the present data and determined \mtop = \gevcc{172.3\pm2.6}~\cite{earlyMETJ}. In the present measurement, we not only use a larger sample but also increase the signal acceptance with changes in the event-selection criteria and improve the sensitivity with a new event-reconstruction method. These changes produce an improvement of about 18\% in statistical precision over the improvement expected from increasing the sample size alone.  We use the full data set of \ppbar collisions collected by the CDF II detector at the Fermilab Tevatron, corresponding to an integrated luminosity of \invfb{8.7}. 

The CDF II detector~\cite{cdfdet} is a general-purpose azimuthally and forward-backward symmetric detector surrounding the colliding beams of the Tevatron \ppbar collider.  A charged-particle tracking system, consisting of an inner silicon microstrip detector and an outer drift chamber,  immersed in a 1.4~T magnetic field, provides accurate vertex and momentum reconstruction.  Electromagnetic and hadronic calorimeters surround the tracking system and measure particle energies.  Drift chambers and scintillators, located outside the calorimeters, detect muon candidates.

The data used in this measurement are collected with a purely calorimetric online selection~(trigger).  Calorimeter energy deposits are clustered into jets using a cone algorithm with an opening angle of $\Delta R \equiv \sqrt{{\Delta\eta}^2+{\Delta\phi}^2}=0.4$~\cite{cordSystem}.  Events are triggered by selecting those containing at least four clusters with $E_{T} > \gev{15}$ and a scalar sum of the $E_T$ of all the clusters greater than \gev{175}.

After the online selection, event observables of physical interest are computed. Jets are reconstructed with the {\sc jetclu}~\cite{jetclu} algorithm using a cone radius of $\Delta R =0.4$.  Jet energies are corrected~\cite{jes} for nonuniformities of the calorimeter response parametrized as a function of $\eta$, energy contributed by multiple \ppbar interactions in the event, and calorimeter nonlinearity. We identify jets originating from the decay of a $b$-quark using the {\sc secvtx} algorithm~\cite{secvtx}.  We require at least one jet to be identified as $b$-quark~($b$-tagging). In order to improve the analysis' sensitivity, we group the candidate events into two samples, one with exclusively single-$b$-tagged events~(1-tag), and the other with events containing two or more $b$-tagged jets~(2-tag). Events are required to have four, five, or six jets with transverse energy $\et > \gev{15}$ and $|\eta| <2.0$. To maintain the event sample independent from those used in other CDF top-quark mass measurements~\cite{cdflj3,cdflj2,cdfdil,cdfall1}, we require events to have no identified electrons or muons with $\pt > \gevc{20}$ and $|\eta| < 1.1$. In order to reject multijet backgrounds from QCD processes, we require the events to have \mets significance (${\mets}^{\text{sig}} \equiv \mets/\sqrt{\sum_{\text{jets}} E_T}$) to be greater than 3~GeV$^{1/2}$, where the sum in the denominator runs over all identified jets in an event. The remaining events have appreciable background from QCD processes due to mismeasurement of jet energies. Because these events sometimes have misalignment between \met and \misspt, the missing transverse-momentum of the event computed using charged tracks~\cite{missptc}, we require $\Delta\phi(\met,\misspt)$, the azimuthal angle between \met and \misspt, to be less than 2.0. 

Background events with $b$-tags arise from QCD multijet events and from electroweak production of $W$-bosons associated with jets. We estimate the background rate using a data-driven method~\cite{earlyMETJ}. This method uses events with exactly three jets, which have negligible ($<$0.1\%) \ttbar component, and employs a per-jet parametrization of the $b$-tagging probability. Due to the presence of \ttbar events in event samples with higher jet multiplicity, we extrapolate the $b$-tagging probability of the three-jet event sample to higher jet multiplicity event samples after iteratively removing the \ttbar content from the samples~\cite{earlyMETJ}. We estimate the background for the 1-tag and 2-tag samples separately. A $b$-tagging correction factor~\cite{cdfall} is applied to account for the dominance of production in pairs for heavy-flavor jets.  In order to improve the signal-to-background ratio in this analysis, an artificial neural network is trained to identify the kinematic and topological characteristics of SM \ttbar events using input variables proposed in Refs.~\cite{METJcross, cdfall}. Compared with our previous work~\cite{earlyMETJ}, we add new input variables, $\Delta\phi(\met,\misspt)$, $\missptc$, and a series of two jets~($2j$) and three jets~($3j$) invariant-masses, $M_{2j}^{\text{min}}$, $M_{2j}^{\text{max}}$, $M_{3j}^{\text{min}}$, and $M_{3j}^{\text{max}}$, where superscripts {\it min} and {\it max} represent the minimum mass and the maximum mass, respectively, among all the possible combinations of 2$j$ or 3$j$. We apply the neural network to all events meeting the above selection criteria. We then define the signal region by requiring a neural network output greater than 0.9 for 1-tag events and 0.8 for 2-tag events, respectively, choosen in order to reject approximately 95\% of background and preserve approximately 80\% of signal.  With this procedure we obtain the estimated numbers of background events in the signal region shown in Table~\ref{nevent}. We also show the expected number of \ttbar signal events, assuming a $\ttbar$ production cross-section of 7.45~pb at \mtop = \gevcc{172.5}~\cite{xseccorr}, together with the number of observed events in the data. Signal events are further separated by the number of jets for reasons explained later.

\begin{table}
\begin{center}
\caption[Number of events]{Numbers of expected signal and estimated background events in the signal region compared to the number of events observed in data.}
\begin{ruledtabular}
\begin{tabular}{ccccc}
\label{nevent}
 &\multicolumn{2}{c}{1-tag} & \multicolumn{2}{c}{2-tag} \\
 \hline
  &4 jets & 5 or 6 jets & 4 jets & 5 or 6 jets \\
\hline
\ttbar &  427 $\pm$ 50 & 801 $\pm$ 70 & 179 $\pm$ 23& 373 $\pm$ 37\\
Background     &  262 $\pm$ 22 & 450 $\pm$ 29 & 43 $\pm$ 11 & 125 $\pm$ 23\\
\hline
\hline
Expected       &  690 $\pm$ 55 & 1251 $\pm$ 76& 222 $\pm$ 26& 498 $\pm$ 44\\
Observed        &   761         &   1341    &       225    & 550 \\
\end{tabular}
\end{ruledtabular}
\end{center}
\end{table}

To distingush between different values of \mtop, we compare the reconstructed top-quark mass distribution from our data to a series of \ttbar signal samples generated by {\sc pythia}~\cite{pythia} with 76 different \mtop values ranging from \gevcc{150} to \gevcc{240}.  Because the jet energy scale~(JES) is one of the dominant systematic uncertainties in the \mtop measurement, we generate a set of samples with JES variations. Data jets in the analysis are corrected by a factor of 1+\djes to account for the scale error in the calorimeter. In the simulation, the value of \djes are varied from $-3.0 \sigma_c$ to $+3.0 \sigma_c$, where $\sigma_c$ is the CDF JES fractional uncertainty~\cite{jes}. 

After event selection, the analysis proceeds in three steps. First we reconstruct two different top-quark masses~(\ma and \mb) using measured jets and \met. We modify the standard $\chi^2$-like kinematic fitter~\cite{cdflj,LJtmass}, which has been used in the lepton+jets channel measurements, for the reconstruction of the lepton+jets with no reconstructed lepton. \ma is the reconstructed top-quark mass from the lowest $\chi^2$ fit between measured jets to partons combinatorics while \mb is taken from the assignment that yields the second lowest $\chi^2$ to increase the statistical power of the measurement. Both \ma and \mb are the senstive variables for \mtop. We also reconstruct the hadronically decaying $W$-boson mass. With the constraint of the well-known $W$-boson mass, this variable can be used to determine the JES calibration {\it in situ} which reduces the dominant uncertainty from the JES. The second step is a likelihood fit of the three variables using simulated signal and background distributions to obtain the measured top-quark mass~($M_t^{\text{meas}}$).  Calibration factors relating this likelihood fit result to \mtop are obtained.  In this process, a three-dimensional kernel density estimation~\cite{cdflj,kde1} is applied to obtain probability density functions~(p.d.f.s) of the signals and background.
Finally, we perform the same likelihood fit to the data and apply the calibration factors to obtain \mtop.

Events used in this measurement have two missing particles, a neutrino and a charged lepton, which are assumed to have two decay products of a $W$-boson. For the \mtop measurement, the reconstruction of $W$-decay particles is not necessary, so these events can be considered as having one missing particle, a $W$-boson that decays leptonically. We then reconstruct  events with a number of constraints that are larger than the number of unknown quantities. We assume that all selected events are lepton+jets \ttbar events with a missing particle, the $W$-boson. Measured four-vectors of jets are corrected for known effects and appropriate resolutions are assigned. The unclustered transverse energy ($\vec{U_T}$)~\cite{cdflj} is estimated as a sum of all transverse energy in the calorimeters that is not associated with one of the selected four jets. The longitudinal momentum of the leptonically-decaying $W$-boson is a free parameter which is effectively determined by the constraints from the  known mass of the $W$-boson and the assumption that $M_{t} = M_{\bar{t}}$, where $M_{t}$ and $M_{\bar{t}}$ are the mass of the top quark and anti-top quark, respectively.  To estimate the reconstructed top-quark mass, \mtr, we define a kinematic $\chi^{2}$ function, 
\begin{eqnarray}
\label{mtchi2}
{\chi}^2=&&{\sum\limits_{i=4 \text{jets}}}\frac{(p^{i,\text{fit}}_T-p^{i,\text{meas}}_T)^2}{{\sigma}_i^2}+\sum\limits_{k = x,y} \frac{(U_{T_k}^{\text{fit}}-U_{T_k}^{\text{meas}})^2}{{\sigma}^2_k} \nonumber \\
&+& \frac{(M_{jj}-M_W)^2}{{\Gamma}^2_W} + \frac{(M_{\text{missing}}-M_W)^2}{{\Gamma}^2_W} \nonumber \\
      &+& \frac{(M_{b,\text{missing}}-\mtr)^2}{{\Gamma}^2_t} + \frac{(M_{bjj}-\mtr)^2}{{\Gamma}^2_t},
\end{eqnarray}
where the value of the free parameter \mtr is determined as the reconstructed top-quark mass value corresponding to the minimum $\chi^2$.  In Eq.~(\ref{mtchi2}), we constrain the four selected jets \pt to their measured values and uncertainties~($\sigma_i$). We also constrain the $x$ and $y$ components of $\vec{U_{T}}$ in the second term which is related to the transverse momentum of the missing $W$-boson.  The third term constrains the dijet mass of the two jets assigned as $W$-decay products to the known $W$-mass within the $W$-boson decay-width. The fourth term constrains the invariant mass of the missing particle~($M_{\text{missing}}$) to the $W$-boson mass. The fifth term constrains the invariant mass of the missing particle and the $b$-quark~(that is regarded as coming from the daughters of the same top-quark decay) to be consistent with the hadronically-decaying top-quark mass within the top-quark decay-width of \gevcc{1.5}. The last term imposes the same constraint on the invariant mass of the two jets regarded as $W$-boson decay products and the $b$-quark that is assigned as coming from the same top-quark decay. 

The event reconstruction described above, using the leading~(highest-\pt) four jets, does not consider the contribution of hadronically-decaying $\tau$ leptons.  However because the $\tau$ lepton can be misidentified as a jet, we also consider five leading jets and assign one of the jets as the misidentified $\tau$ lepton. We perform the $\chi^2$ fit in Eq.~(\ref{mtchi2}) for each possible jet-to-parton assignment.  Assuming that the leading five jets in any event come from the four final quarks and one hadronically-decaying $\tau$ lepton, there are 24 and 6 possible assignments of jets to quarks or $\tau$-lepton for 1-tag and 2-tag, respectively. In the case of four-jet events, we assume that the four jets come from the four quarks. This makes 6 and 2 possible combinations in 1-tag and 2-tag, respectively. The $\chi^2$ minimization is performed for each jet-to-quark or jet-to-$\tau$ assignment, and the first variable \mtr is taken from the assignment that yields the lowest $\chi^2$. Due to the differing number of assignments between events with four jets and those with five or six, the resolution of \mtr is different. We therefore separate the candidate events accordingly. 

In order to extract more statistical information from each event, we add a second variable, \mb, the reconstructed top-quark mass that corresponds to the second-lowest $\chi^2$~\cite{cdflj2} in the jet-to-quark and jet-to-$\tau$ combinatorics.  Studies based on MC samples show that \ma and \mb have better sensitivity to the input top-quark masses of the samples than the two estimators used in a previous analysis~\cite{earlyMETJ}.

The third variable, \mjj, defined as the invariant mass of the two jets from the hadronically-decaying $W$-boson, serves as an {\it in situ} constraint on the JES through the likelihood fit. We calculate \mjj from the two non-$b$-tagged jets. If  more than two non $b$-tagged jets are present, we use the closest value to the world average $W$-boson mass, \gevcc{80.40}~\cite{wmass}, from all possible combinations. 

By accounting for the correlations between \ma, \mb, and \mjj, we reconstruct three-dimensional p.d.f.s of signals and background for the likelihood fit procedure. First, we estimate p.d.f.s for the observables from the above-mentioned {\sc pythia} \ttbar samples at discrete values of \mtop from \gevcc{150} to \gevcc{240} and \djes from $-3.0 \sigma_c$ to $+3.0 \sigma_c$. Background p.d.f.s are estimated for discrete \djes. We interpolate the MC distributions to find p.d.f.s for arbitrary values of \mtop and \djes using the local polynomial smoothing method~\cite{lps}. Then, we fit the signal and background p.d.f.s to the unbinned distributions observed in the data. Separate likelihoods are built for the four subsamples, and the overall likelihood is obtained by multiplying them together. References~\cite{cdflj,like1} provide detailed information about this technique.

The mass fitting procedure is tested with pseudoexperiments for a set of MC-simulated \ttbar samples with 14 different \mtop values ranging from \gevcc{159} to \gevcc{185}. For each pseudoexperiment, we draw the number of background events from a Poisson distribution with a mean equal to the estimated total number of background events in the sample, and the number of signal events from a Poisson distribution with a mean equal to the expected number of signal events normalized to a \ttbar production cross-section of 7.45 pb. The mean value of the distributions of the mass residual~(the deviation from the input top-quark mass) for simulated experiments is corrected to be zero, and the correction from linear regression analysis is $M_{t}^{\text{corr}} = 1.066 \times M_t^{\text{meas}}- 11.46$ \gevccnoarg, where $M_t^{\text{meas}}$ is the raw value from the likelihood fit and $M_t^{\text{corr}}$ is the corrected value of the measurement.  The width of the pull is consistent with unity after the correction.  We also test the mass fit results using different values of \djes between $-1.0 \sigma_c$ and $+1.0 \sigma_c$ with three different \mtop points, 168, 173, and 178~\gevccnoarg. With the correction discussed above, the residuals of \mtop from different \djes values are consistent with zero in case of \mtop=\gevcc{168} and \gevcc{173}. However, the pseudoexperiments corresponding to a top-quark mass of \gevcc{178} show a  \gevcc{0.42} difference between $-1.0 \sigma_c$ and $+1.0 \sigma_c$. We take the half difference (\gevcc{0.21}) as the systematic uncertainty on the calibration.

We examine the effect of various sources of systematic uncertainties by comparing the results of pseudoexperiments in which we vary relevant parameters within their uncertainties. One of the leading sources of systematic uncertainty is the residual JES~\cite{jes,cdflj}. We vary the JES components within their uncertainties in the MC-simulated signal events and interpret the shifts in the returned top-quark mass as uncertainties. The $b$-jet energy scale systematic uncertainty that arises from the modeling of $b$ fragmentation, $b$-hadron branching fractions, and calorimeter response captures the additional uncertainties not included in the light-quark-jet energy scale~\cite{cdflj}. The uncertainty arising from the choice of MC generator is estimated by comparing results from MC samples generated with {\sc pythia} and {\sc herwig}~\cite{herwig}. We estimate the systematic uncertainty due to imperfect modeling of initial-state gluon radiation and final-state gluon radiation by varying the amounts of initial- and final-state radiations in simulated events~\cite{LJtmass}. We estimate the systematic uncertainty due to parton distribution functions~(PDF's) of the proton by varying the independent eigenvectors of the CTEQ6M~\cite{cteq6} PDF's, varying the QCD scale $\Lambda_{\text{QCD}}$~(228 MeV vs. 300 MeV), and comparing CTEQ5M~\cite{cteq5} with MRST72~\cite{mrst72} PDF's. To estimate the systematic uncertainty associated with uncertainties in the top-quark production mechanism, we vary the fraction of the top quarks produced by gluon-gluon annihilation from the default 6$\%$ to 20$\%$, corresponding to a one-standard-deviation upper bound on the gluon fusion fraction~\cite{gg}.  The background systematic uncertainty accounts for the variation of the background originating from the uncertainty on the per-jet $b$-tagging probability. It includes not only shape change of the reconstructed variables but also background normalization.  The trigger efficiency is estimated using a combination of MC and data~\cite{trigco}.  We evaluate the uncertainty propagated from the corrections of the trigger efficiency in the signal MC samples. 
We also estimate an uncertainty due to the effect of multiple hadron interactions, including its dependence on the instantaneous luminosity profile of the data. The color reconnection (CR) systematic uncertainty~\cite{cr2} is evaluated using MC samples generated with and without CR effects adopting different configurations of {\sc pythia}~\cite{mt2}. Table~\ref{systtable} summarizes all systematic uncertainties, which summed in quadrature, total to \gevcc{0.87}. 

\begin{table}
\begin{center}
\caption[Summary of systematics]{Systematic uncertainties on the top-quark mass measurement.}
\begin{ruledtabular}
\begin{tabular}{cc}
\label{systtable}
Source & Uncertainty~(\gevccnoarg) \\
\hline
Residual jet-energy scale  & 0.44 \\
MC generator               & 0.36 \\
Color reconnection     &0.28 \\
$gg$ fraction          &0.27 \\
Radiation              &0.28 \\   
PDFs                   &0.16 \\
$b$-jet energy scale   &0.19 \\
Background             &0.15 \\
Calibration            &0.21 \\
Multiple hadron interaction   &0.18 \\                       
Trigger modeling        & 0.13 \\
        \hline                      
Total systematic uncertainty       &0.87\\
\end{tabular}
\end{ruledtabular}
\end{center}
\end{table}

By applying a likelihood fit to the data using the three variables described above and the corrections obtained from the simulated experiments, the top-quark mass is measured to be 
\begin{eqnarray}
	\mtop & = & \gevcc{\measStatSyst{173.93}{1.64}{0.87}} \nonumber \\
	& = & \gevcc{\measErr{173.93}{1.85}}.
\end{eqnarray}

Figures~\ref{fig:pdf1d4jets} and~\ref{fig:pdf1d} show the observed distributions of the variables used for the \mtop measurement overlaid with density estimates using \ttbar signal events with \mtop = \gevcc{173.5} and the background model. Graphs are presented for events with four jets, and five or six jets, respectively. 

\begin{cfigure1c}
\begin{tabular}{ccc}
\includegraphics[width=0.33\textwidth]{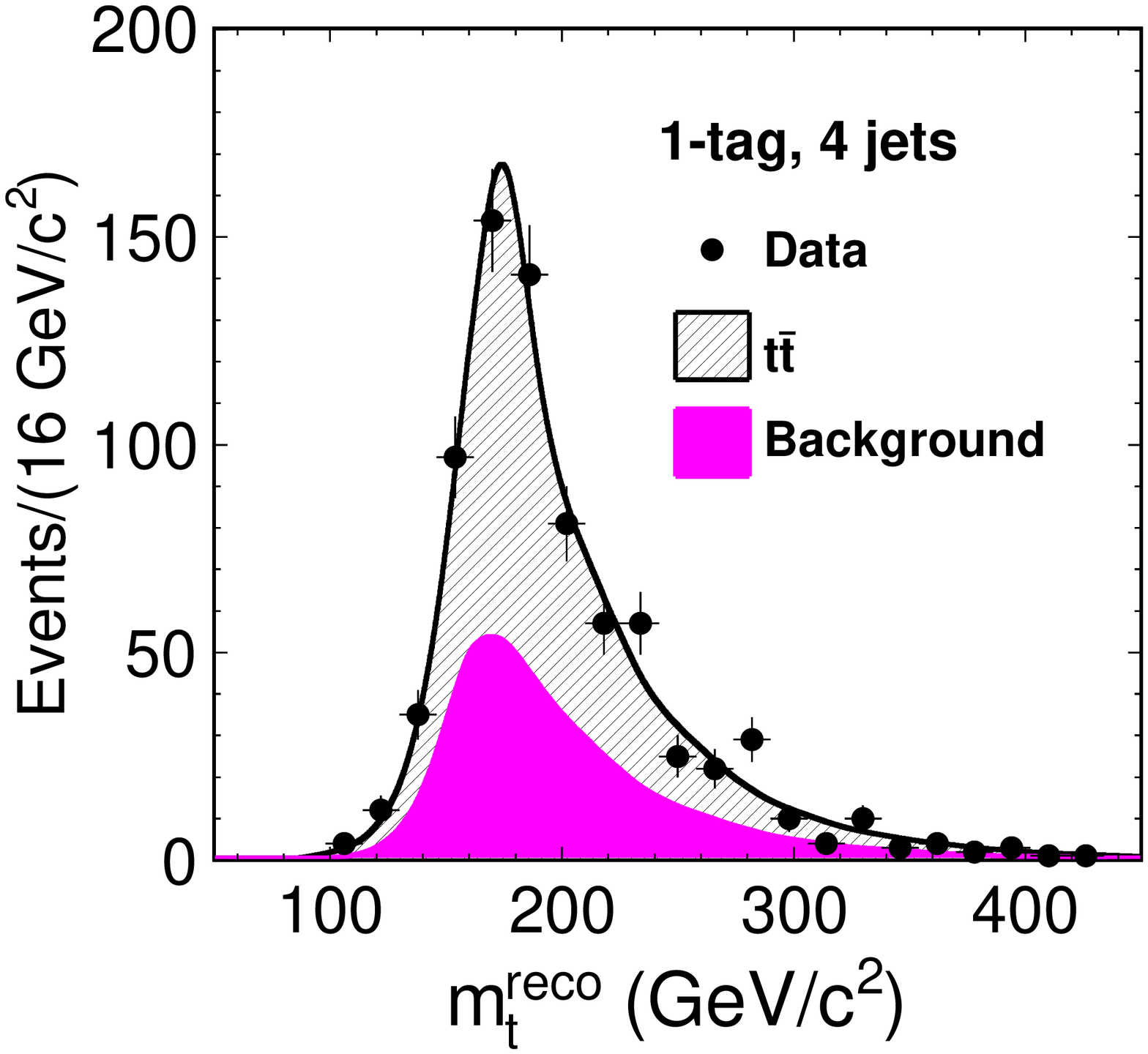} &
\includegraphics[width=0.33\textwidth]{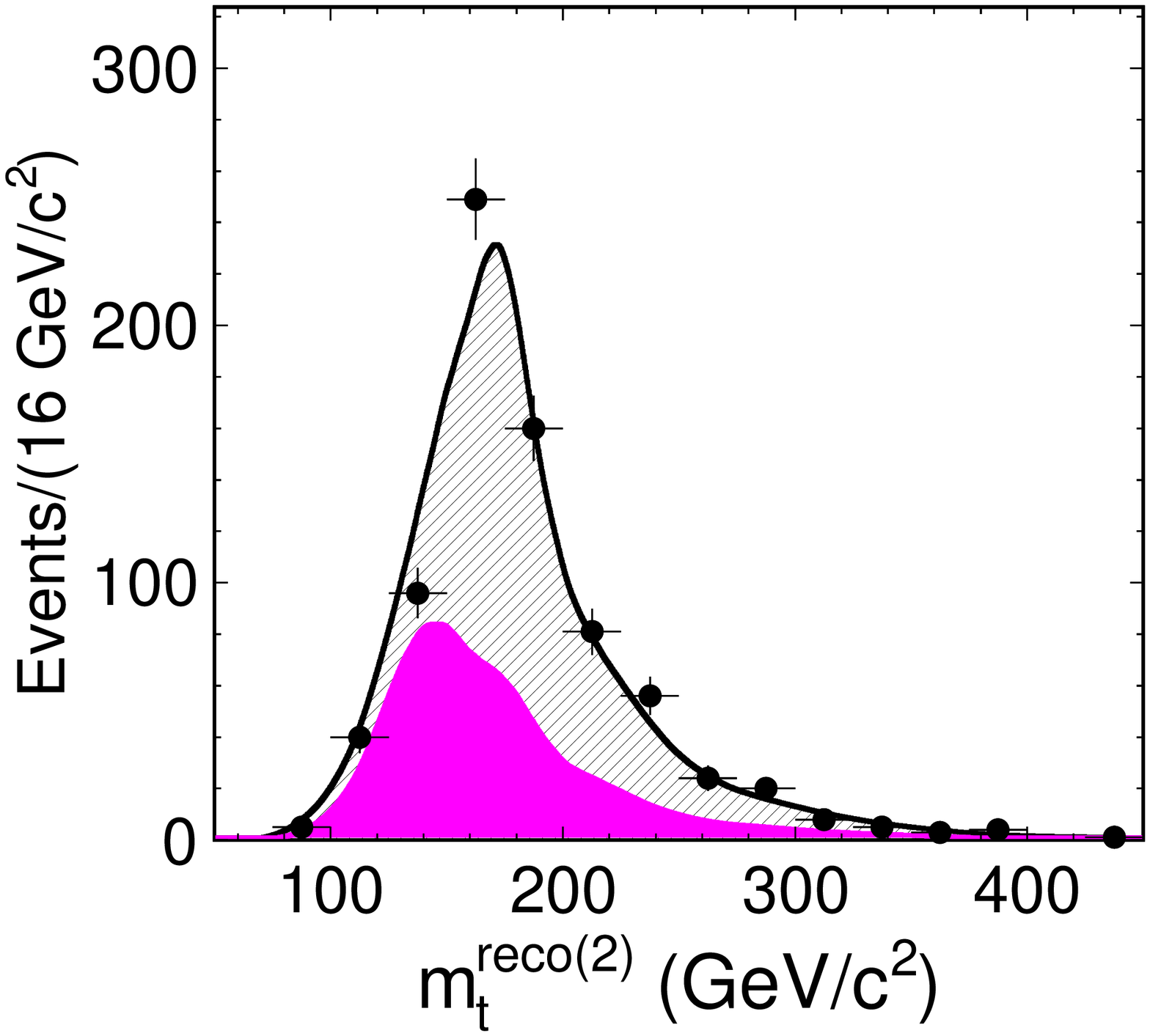}&
\includegraphics[width=0.33\textwidth]{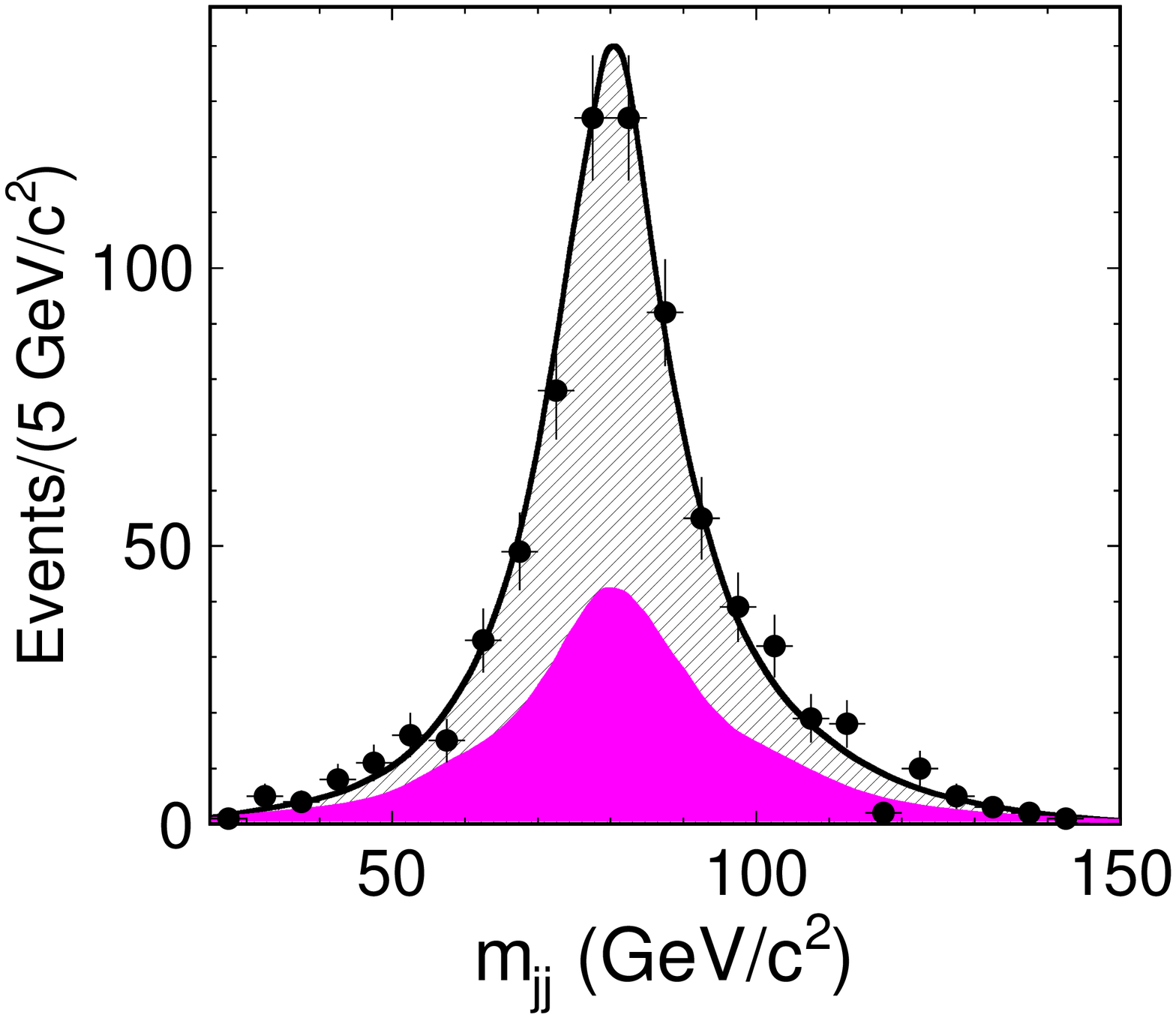} \\
\includegraphics[width=0.33\textwidth]{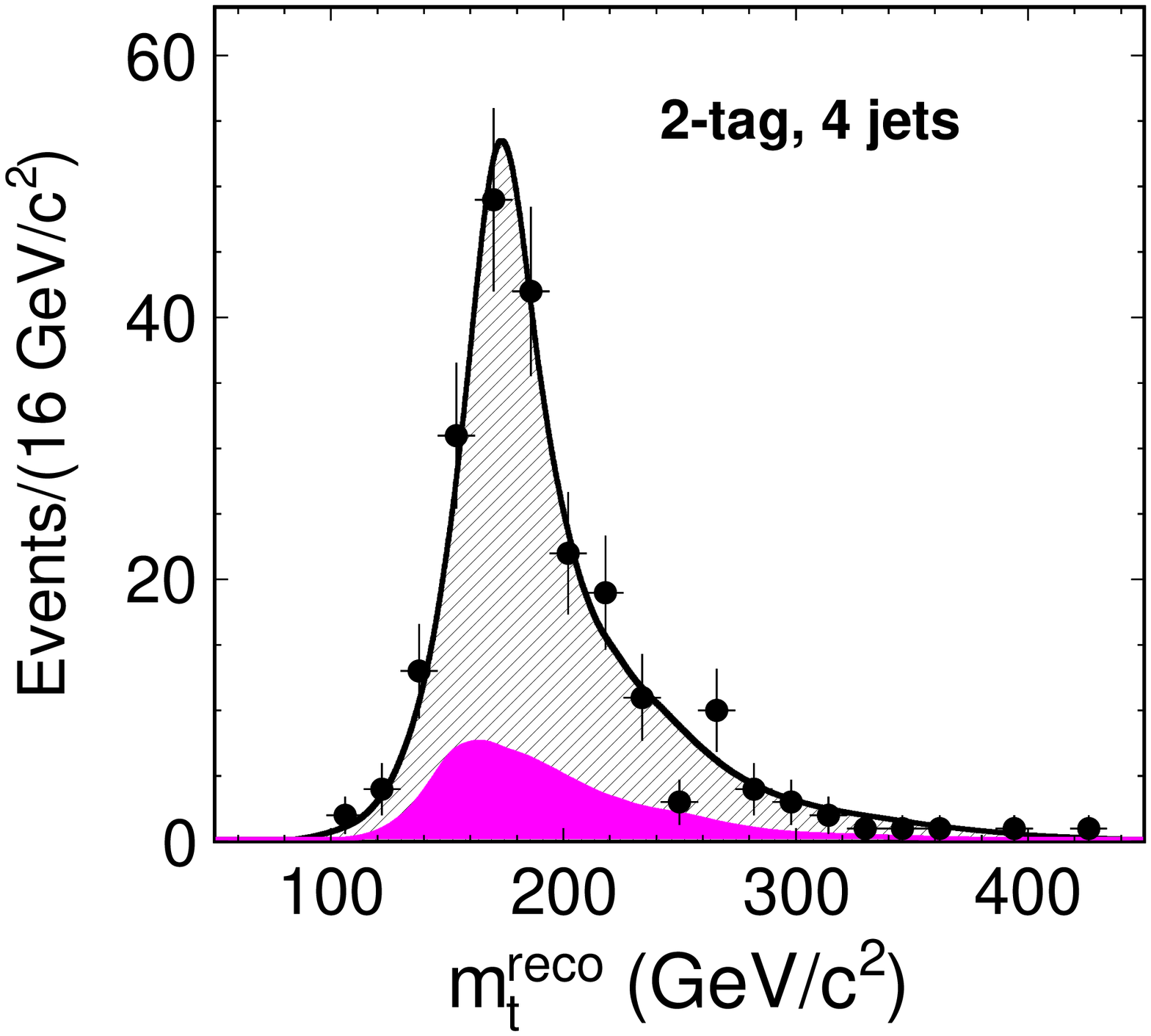} &
\includegraphics[width=0.33\textwidth]{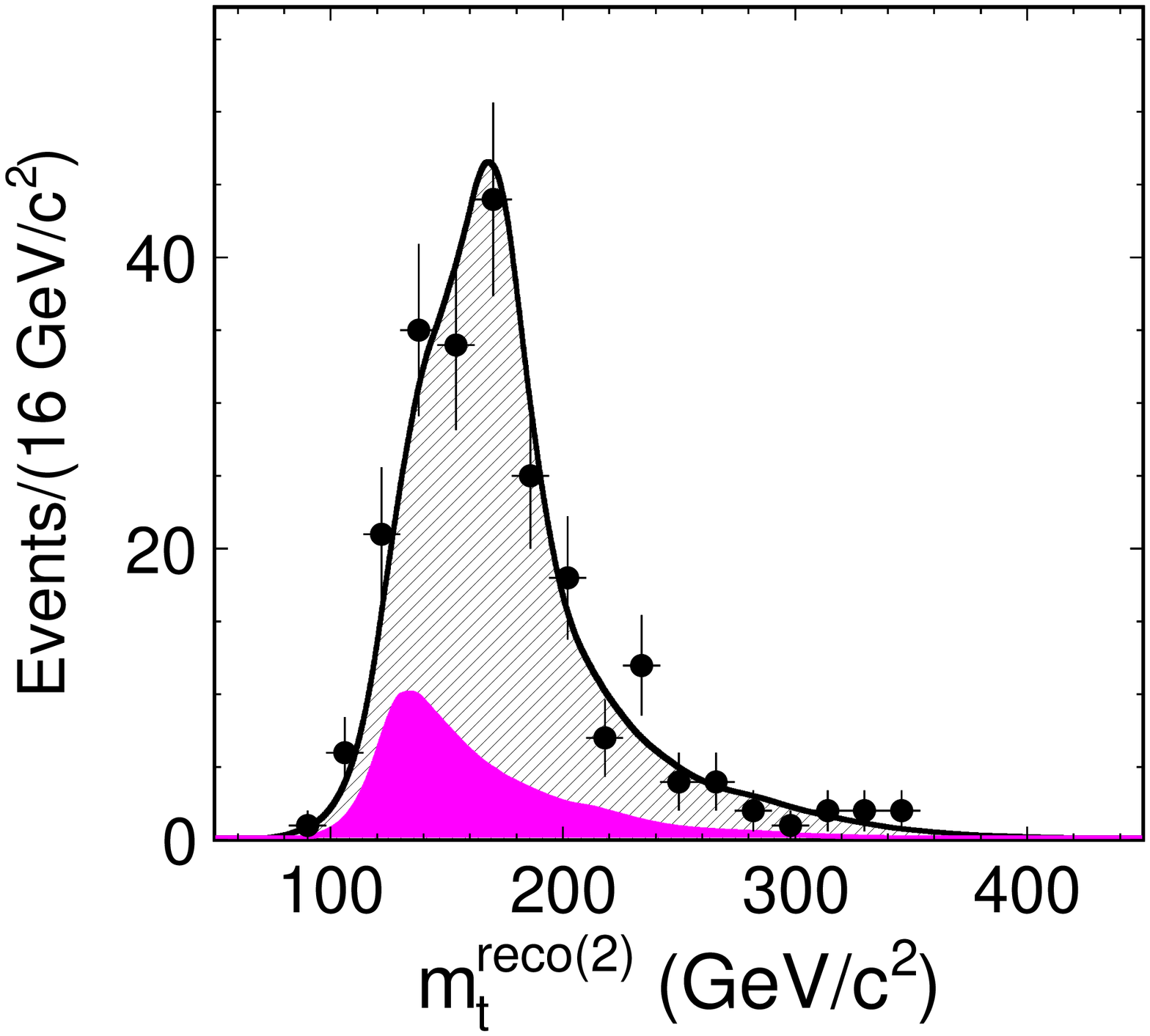}&
\includegraphics[width=0.33\textwidth]{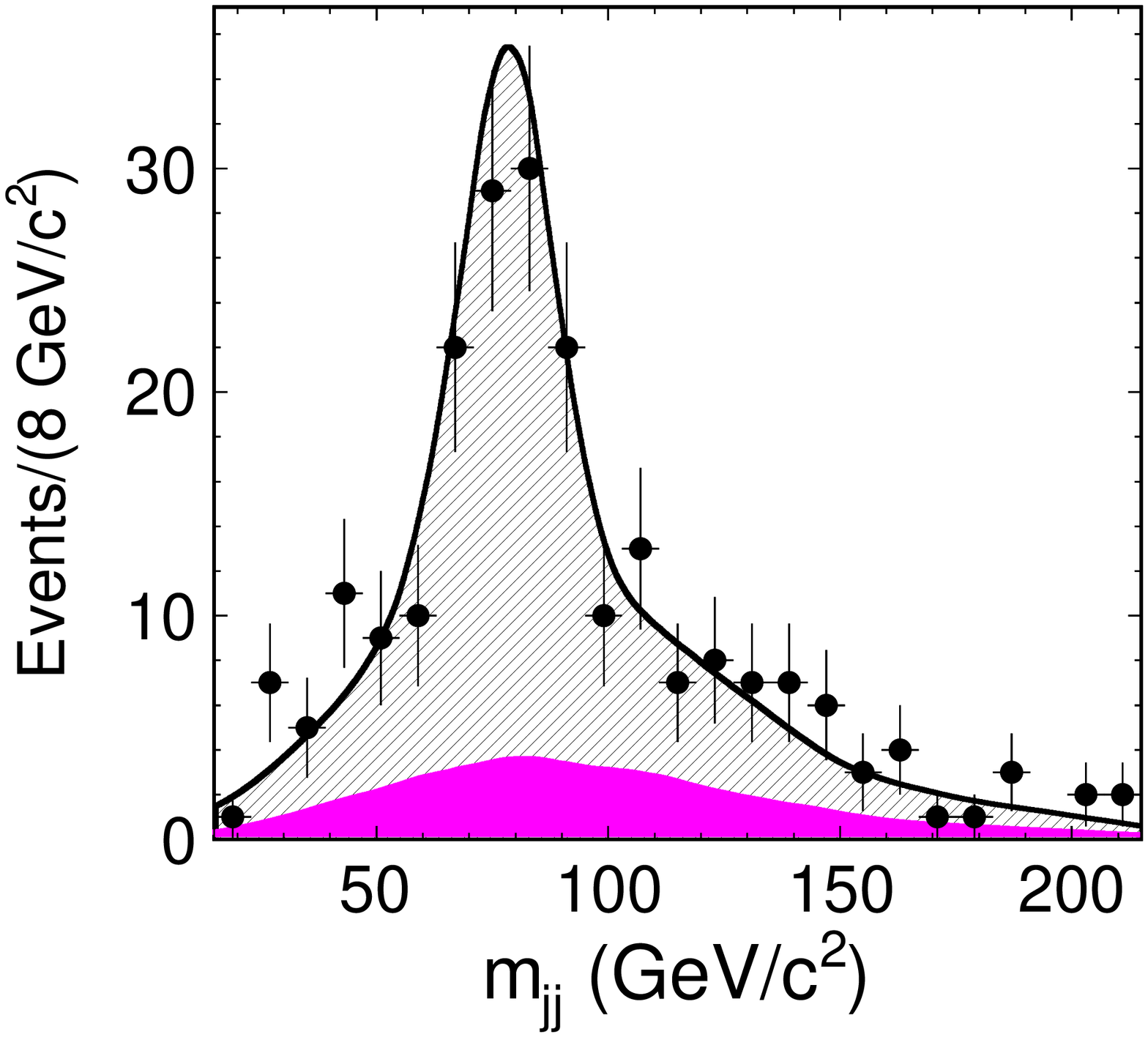} \\
\end{tabular}
\caption[pdff]{Distribution of three variables \ma, \mb, and \mjj for events with four jets, from data~(points), overlaid with their corresponding one-dimensional p.d.f.s from signal MC sample~(\mtop = \gevcc{173.5}, hashed area) plus the estimated background~(filled area). The 1-tag~(top) and 2-tag~(bottom) distributions are separately shown. 
}
\label{fig:pdf1d4jets}
\end{cfigure1c}

\begin{cfigure1c}
\begin{tabular}{ccc}
\includegraphics[width=0.33\textwidth]{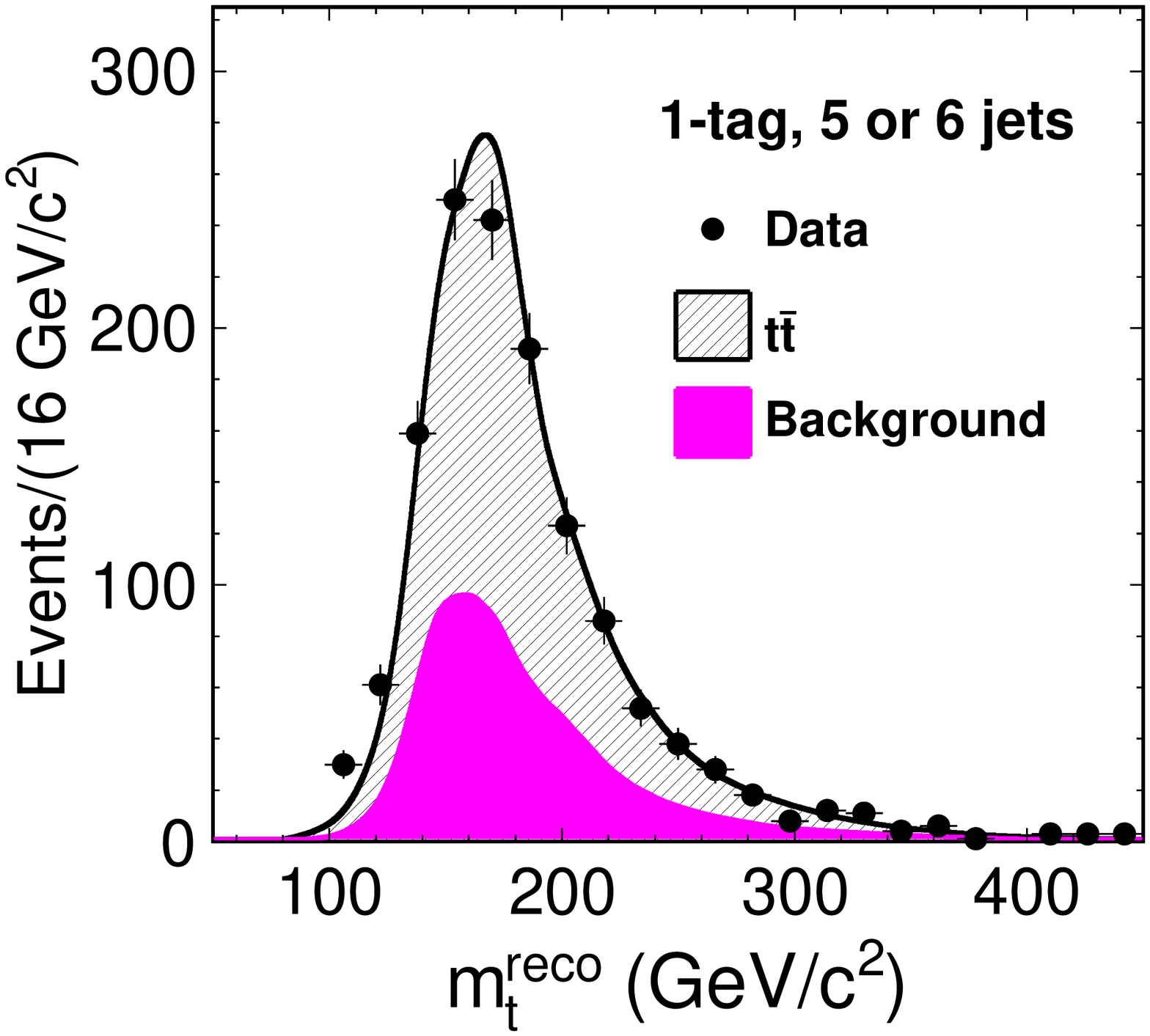} &
\includegraphics[width=0.33\textwidth]{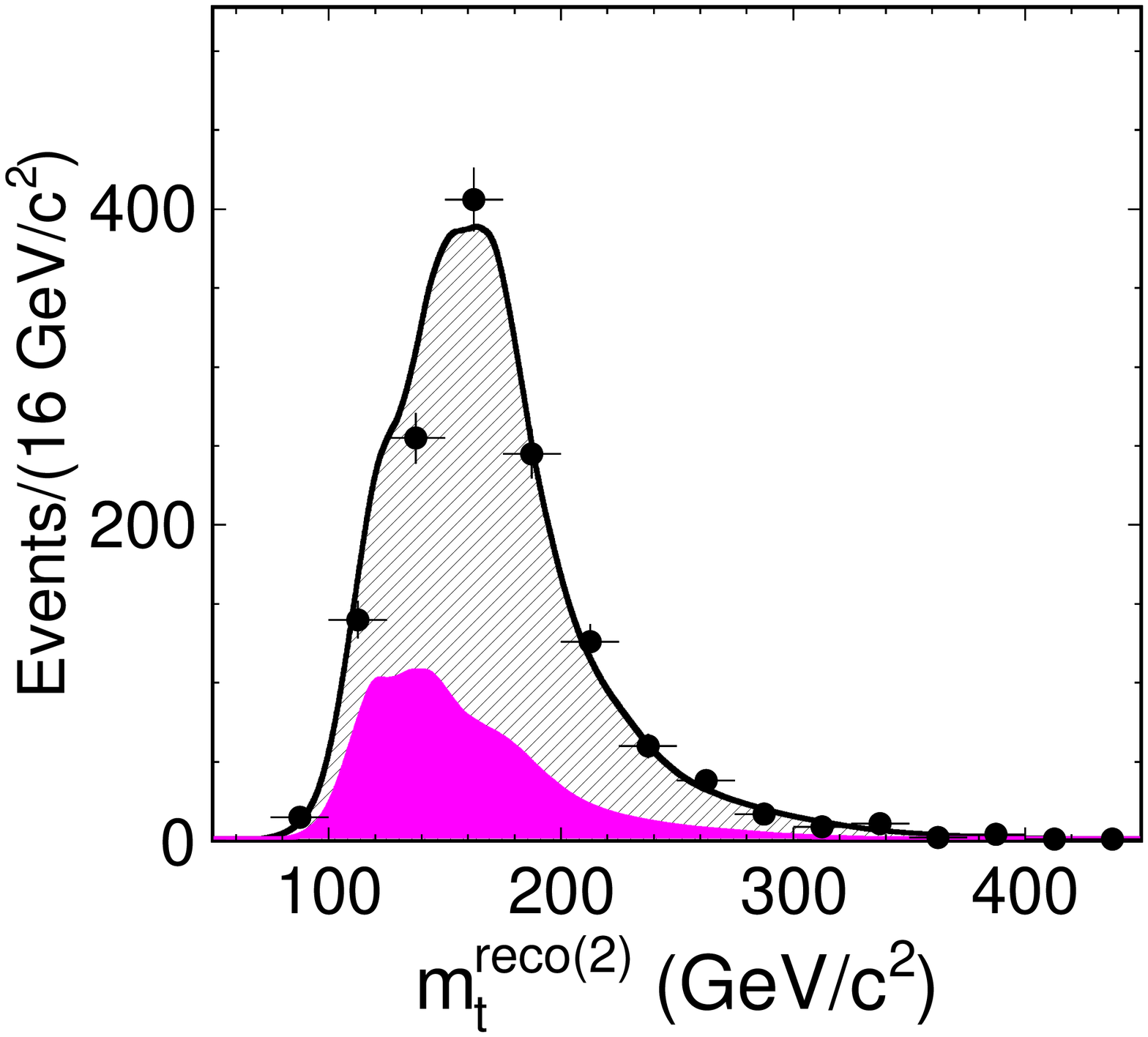}&
\includegraphics[width=0.33\textwidth]{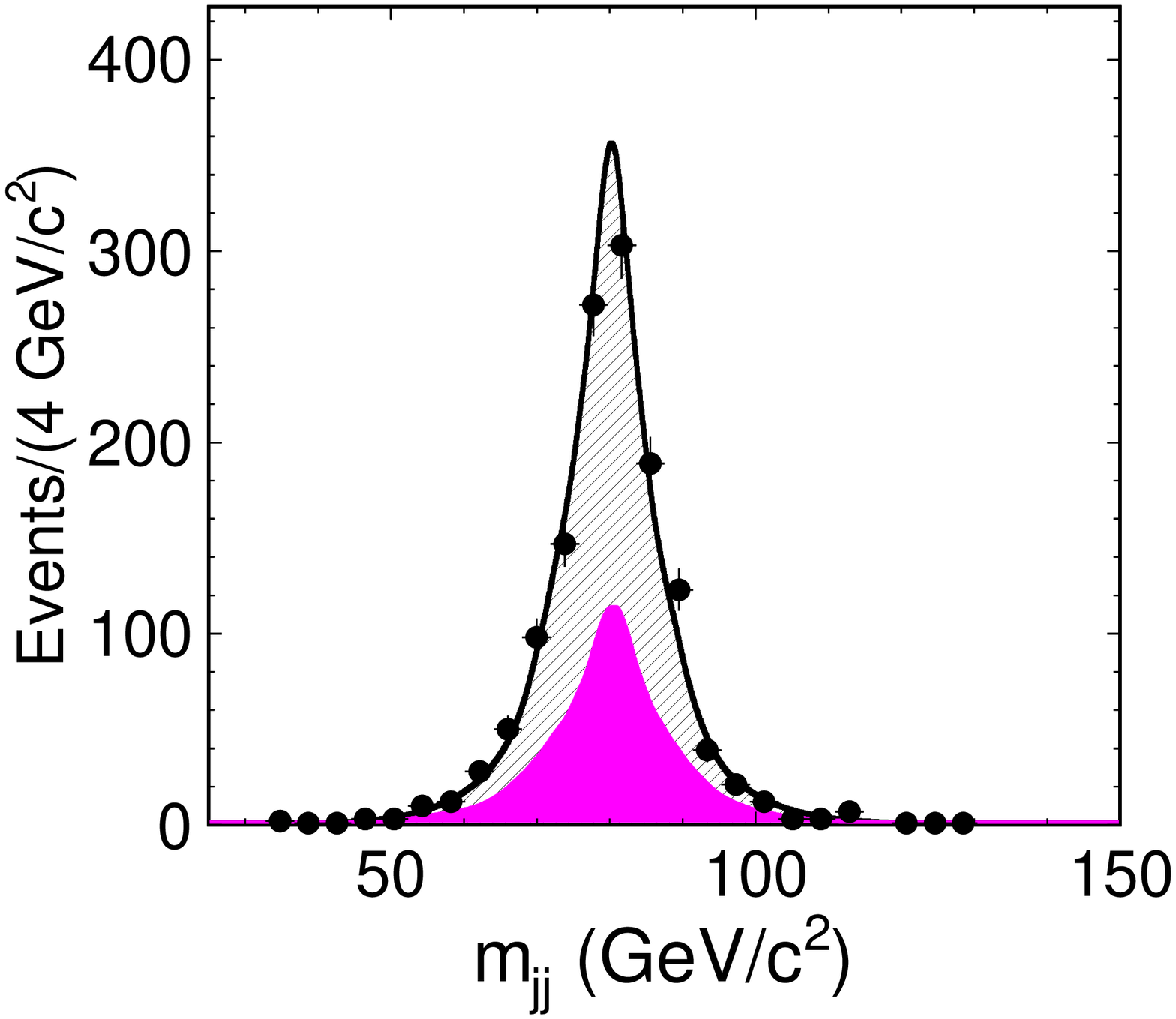} \\
\includegraphics[width=0.33\textwidth]{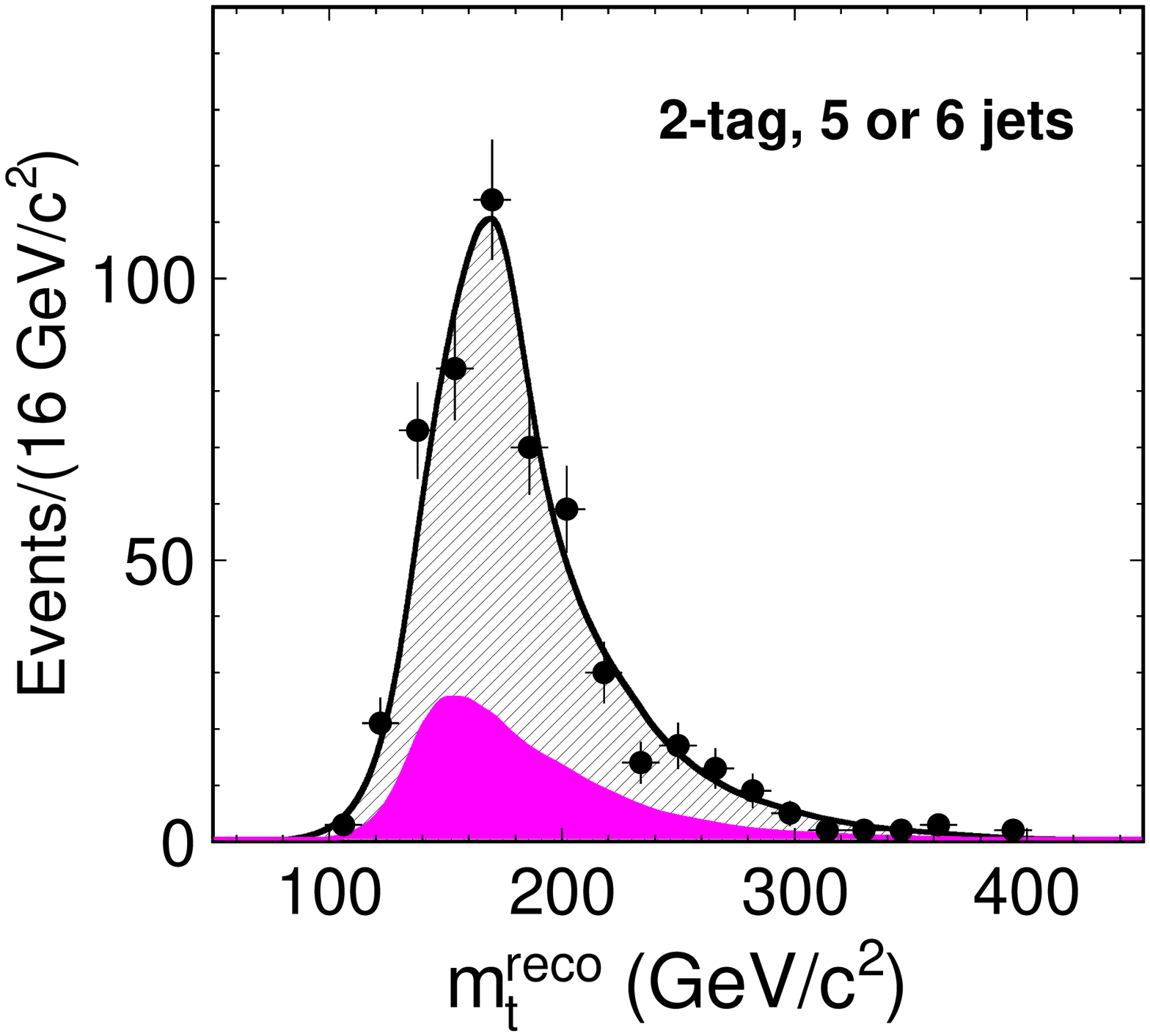} &
\includegraphics[width=0.33\textwidth]{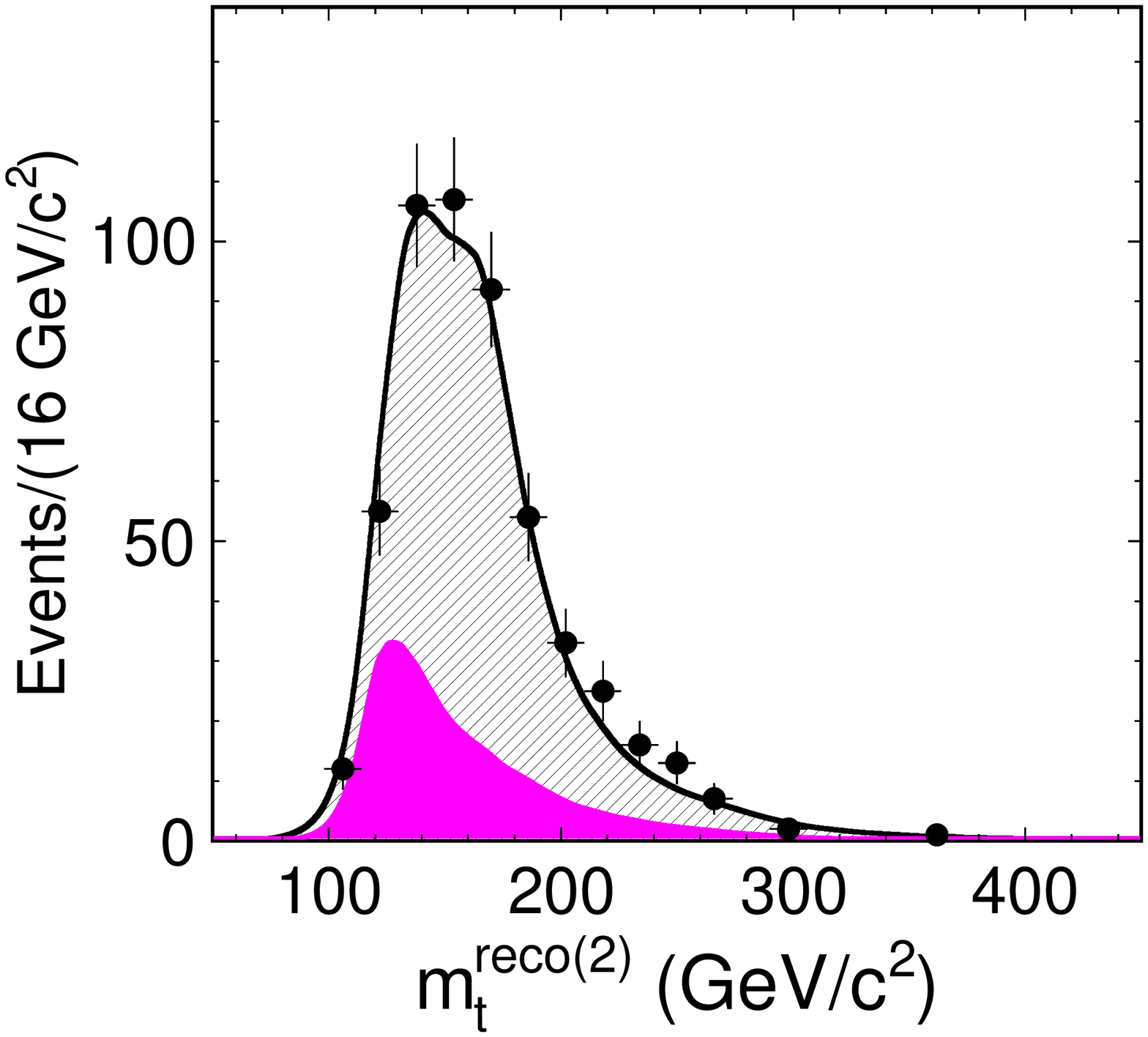} &
\includegraphics[width=0.33\textwidth]{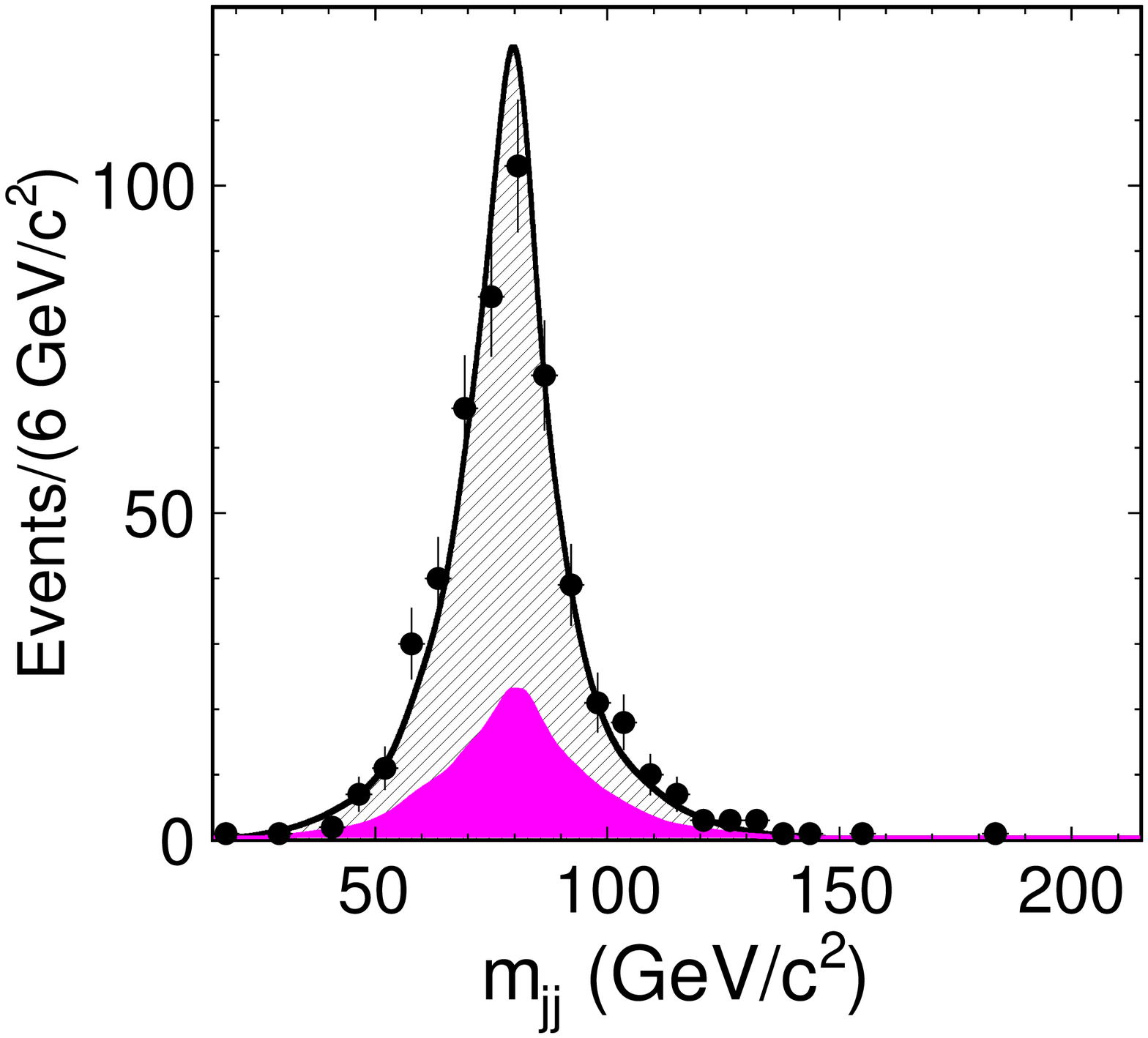} \\
\end{tabular}
\caption[pdff]{Distribution of three variables \mjj, \ma, and \mb for events with five or six jets, from data~(points), overlaid with their corresponding one-dimensional p.d.f.s from signal MC sample~(\mtop = \gevcc{173.5}, hashed area) plus the estimated background~(filled area). The 1-tag~(top) and 2-tag~(bottom) distributions are separately shown. 
}
\label{fig:pdf1d}
\end{cfigure1c}

In conclusion, we perform a measurement of the top-quark mass in events with jets and large \mets in data corresponding to an integrated luminosity of  \invfb{8.7} collected by the CDF experiment. The data sample is chosen in such a way as to be statistically independent from samples used in other CDF top-quark mass measurements, apart from the earlier version of this work~\cite{earlyMETJ}. The result, $\mtop = \gevcc{\measErr{173.93}{1.85}}$, is a considerable improvement on the previous measurement with the same event signature, and is in agreement with the recent published Tevatron average of $\mtop = \gevcc{\measErr{173.18}{0.94}}$~\cite{tevaverage}. This result is included in the most recent preliminary Tevatron average with approximately 12\% weight while a measurement using the standard lepton+jets channel with same data set contributes approximately 62\% weight~\cite{tevaverage1}. The present result is the most precise top-quark mass measurement to date in this event topology.

\begin{acknowledgments}
	We thank the Fermilab staff and the technical staffs of the participating institutions for their vital contributions. This work was supported by the U.S. Department of Energy and National Science Foundation; the Italian Istituto Nazionale di Fisica Nucleare; the Ministry of Education, Culture, Sports, Science and Technology of Japan; the Natural Sciences and Engineering Research Council of Canada; the National Science Council of the Republic of China; the Swiss National Science Foundation; the A.P. Sloan Foundation; the Bundesministerium f\"ur Bildung und Forschung, Germany; the Korean World Class University Program, the National Research Foundation of Korea; the Science and Technology Facilities Council and the Royal Society, UK; the Russian Foundation for Basic Research; the Ministerio de Ciencia e Innovaci\'{o}n, and Programa Consolider-Ingenio 2010, Spain; the Slovak R\&D Agency; the Academy of Finland; the Australian Research Council (ARC); and the EU community Marie Curie Fellowship contract 302103. 
\end{acknowledgments}

\end{document}

%% file: prdDefs.tex
\newenvironment{cfigure1c}[1][tbp]{\begin{figure*}[#1]\centering}{\end{figure*}}

\newcommand{\note}[1]{\xspace}
\newcommand{\robustnote}[1]{\xspace}
\begin{printrobustnotes}
\renewcommand{\robustnote}[1]{\textbf{\textit{#1}}\xspace}
\end{printrobustnotes}
\begin{printallnotes}
\renewcommand{\note}[1]{\textbf{\textit{#1}}\xspace}
\renewcommand{\robustnote}[1]{\textbf{\textit{#1}}\xspace}
\end{printallnotes}
\newcommand{\ppbar}  {\ensuremath{p\bar{p}}\xspace}
\newcommand{\ttbar}  {\ensuremath{t\bar{t}}\xspace}

\newcommand{\et}     {\ensuremath{E_{T}}\xspace}

\newcommand{\pt}     {\ensuremath{p_{T}}\xspace}

\newcommand{\met}    {\ensuremath{\vec{\text{\raisebox{.3ex}{$\not$}}\et}}\xspace}
\newcommand{\mets}    {\ensuremath{\text{\raisebox{.3ex}{$\not$}}\et}\xspace}

\newcommand{\misspt} {\ensuremath{\vec{\text{\raisebox{.3ex}{$\not$}}\pt}}\xspace}
\newcommand{\missptc} {\ensuremath{\text{\raisebox{.3ex}{$\not$}}\pt}\xspace}
\newcommand{\ma}    {\ensuremath{m_{t}^{\text{reco}}}\xspace}
\newcommand{\mb}    {\ensuremath{m_{t}^{\text{reco(2)}}}\xspace}

\newcommand{\mtop}    {\ensuremath{{M}_{\text{top}}}\xspace}

\newcommand{\mtr}     {\ensuremath{m_{t}^{\text{reco}}}\xspace}

\newcommand{\mjj}    {\ensuremath{m_{jj}}\xspace}

\newcommand{\djes}   {\ensuremath{\Delta_{\mathrm{JES}}}\xspace}

\newcommand{\genunit}[2]{\ensuremath{#1~\mathrm{#2}}\xspace}

\newcommand{\gev}[1]    {\genunit{#1}{GeV}}

\newcommand{\gevc}[1]   {\ensuremath{#1~\mathrm{GeV}/c}}
\newcommand{\gevcc}[1]  {\ensuremath{#1~\mathrm{GeV}/c^{2}}}

\newcommand{\invfb}[1]  {\ensuremath{#1~\mathrm{fb}^{-1}}}

\newcommand{\gevccnoarg}{\ensuremath{\mathrm{GeV}/c^{2}}\xspace}

\newcommand{\measErr}[2]{\ensuremath{#1 \pm #2}\xspace}

\newcommand{\measStatSyst}[3]{\ensuremath{#1 \pm #2~(\textrm{stat}) \pm #3~(\textrm{syst})}\xspace}

%% file: November2012_Authors.tex
\affiliation{Institute of Physics, Academia Sinica, Taipei, Taiwan 11529, Republic of China}
\affiliation{Argonne National Laboratory, Argonne, Illinois 60439, USA}
\affiliation{University of Athens, 157 71 Athens, Greece}
\affiliation{Institut de Fisica d'Altes Energies, ICREA, Universitat Autonoma de Barcelona, E-08193, Bellaterra (Barcelona), Spain}
\affiliation{Baylor University, Waco, Texas 76798, USA}
\affiliation{Istituto Nazionale di Fisica Nucleare Bologna, $^{ee}$University of Bologna, I-40127 Bologna, Italy}
\affiliation{University of California, Davis, Davis, California 95616, USA}
\affiliation{University of California, Los Angeles, Los Angeles, California 90024, USA}
\affiliation{Instituto de Fisica de Cantabria, CSIC-University of Cantabria, 39005 Santander, Spain}
\affiliation{Carnegie Mellon University, Pittsburgh, Pennsylvania 15213, USA}
\affiliation{Enrico Fermi Institute, University of Chicago, Chicago, Illinois 60637, USA}
\affiliation{Comenius University, 842 48 Bratislava, Slovakia; Institute of Experimental Physics, 040 01 Kosice, Slovakia}
\affiliation{Joint Institute for Nuclear Research, RU-141980 Dubna, Russia}
\affiliation{Duke University, Durham, North Carolina 27708, USA}
\affiliation{Fermi National Accelerator Laboratory, Batavia, Illinois 60510, USA}
\affiliation{University of Florida, Gainesville, Florida 32611, USA}
\affiliation{Laboratori Nazionali di Frascati, Istituto Nazionale di Fisica Nucleare, I-00044 Frascati, Italy}
\affiliation{University of Geneva, CH-1211 Geneva 4, Switzerland}
\affiliation{Glasgow University, Glasgow G12 8QQ, United Kingdom}
\affiliation{Harvard University, Cambridge, Massachusetts 02138, USA}
\affiliation{Division of High Energy Physics, Department of Physics, University of Helsinki and Helsinki Institute of Physics, FIN-00014, Helsinki, Finland}
\affiliation{University of Illinois, Urbana, Illinois 61801, USA}
\affiliation{The Johns Hopkins University, Baltimore, Maryland 21218, USA}
\affiliation{Institut f\"{u}r Experimentelle Kernphysik, Karlsruhe Institute of Technology, D-76131 Karlsruhe, Germany}
\affiliation{Center for High Energy Physics: Kyungpook National University, Daegu 702-701, Korea; Seoul National University, Seoul 151-742, Korea; Sungkyunkwan University, Suwon 440-746, Korea; Korea Institute of Science and Technology Information, Daejeon 305-806, Korea; Chonnam National University, Gwangju 500-757, Korea; Chonbuk National University, Jeonju 561-756, Korea; Ewha Womans University, Seoul, 120-750, Korea}
\affiliation{Ernest Orlando Lawrence Berkeley National Laboratory, Berkeley, California 94720, USA}
\affiliation{University of Liverpool, Liverpool L69 7ZE, United Kingdom}
\affiliation{University College London, London WC1E 6BT, United Kingdom}
\affiliation{Centro de Investigaciones Energeticas Medioambientales y Tecnologicas, E-28040 Madrid, Spain}
\affiliation{Massachusetts Institute of Technology, Cambridge, Massachusetts 02139, USA}
\affiliation{Institute of Particle Physics: McGill University, Montr\'{e}al, Qu\'{e}bec H3A~2T8, Canada; Simon Fraser University, Burnaby, British Columbia V5A~1S6, Canada; University of Toronto, Toronto, Ontario M5S~1A7, Canada; and TRIUMF, Vancouver, British Columbia V6T~2A3, Canada}
\affiliation{University of Michigan, Ann Arbor, Michigan 48109, USA}
\affiliation{Michigan State University, East Lansing, Michigan 48824, USA}
\affiliation{Institution for Theoretical and Experimental Physics, ITEP, Moscow 117259, Russia}
\affiliation{University of New Mexico, Albuquerque, New Mexico 87131, USA}
\affiliation{The Ohio State University, Columbus, Ohio 43210, USA}
\affiliation{Okayama University, Okayama 700-8530, Japan}
\affiliation{Osaka City University, Osaka 588, Japan}
\affiliation{University of Oxford, Oxford OX1 3RH, United Kingdom}
\affiliation{Istituto Nazionale di Fisica Nucleare, Sezione di Padova-Trento, $^{ff}$University of Padova, I-35131 Padova, Italy}
\affiliation{University of Pennsylvania, Philadelphia, Pennsylvania 19104, USA}
\affiliation{Istituto Nazionale di Fisica Nucleare Pisa, $^{gg}$University of Pisa, $^{hh}$University of Siena and $^{ii}$Scuola Normale Superiore, I-56127 Pisa, Italy, $^{mm}$INFN Pavia and University of Pavia, I-27100 Pavia, Italy}
\affiliation{University of Pittsburgh, Pittsburgh, Pennsylvania 15260, USA}
\affiliation{Purdue University, West Lafayette, Indiana 47907, USA}
\affiliation{University of Rochester, Rochester, New York 14627, USA}
\affiliation{The Rockefeller University, New York, New York 10065, USA}
\affiliation{Istituto Nazionale di Fisica Nucleare, Sezione di Roma 1, $^{jj}$Sapienza Universit\`{a} di Roma, I-00185 Roma, Italy}
\affiliation{Mitchell Institute for Fundamental Physics and Astronomy, Texas A\&M University, College Station, Texas 77843, USA}
\affiliation{Istituto Nazionale di Fisica Nucleare Trieste/Udine; $^{nn}$University of Trieste, I-34127 Trieste, Italy; $^{kk}$University of Udine, I-33100 Udine, Italy}
\affiliation{University of Tsukuba, Tsukuba, Ibaraki 305, Japan}
\affiliation{Tufts University, Medford, Massachusetts 02155, USA}
\affiliation{University of Virginia, Charlottesville, Virginia 22906, USA}
\affiliation{Waseda University, Tokyo 169, Japan}
\affiliation{Wayne State University, Detroit, Michigan 48201, USA}
\affiliation{University of Wisconsin, Madison, Wisconsin 53706, USA}
\affiliation{Yale University, New Haven, Connecticut 06520, USA}

\author{T.~Aaltonen}
\affiliation{Division of High Energy Physics, Department of Physics, University of Helsinki and Helsinki Institute of Physics, FIN-00014, Helsinki, Finland}
\author{S.~Amerio}
\affiliation{Istituto Nazionale di Fisica Nucleare, Sezione di Padova-Trento, $^{ff}$University of Padova, I-35131 Padova, Italy}
\author{D.~Amidei}
\affiliation{University of Michigan, Ann Arbor, Michigan 48109, USA}
\author{A.~Anastassov$^x$}
\affiliation{Fermi National Accelerator Laboratory, Batavia, Illinois 60510, USA}
\author{A.~Annovi}
\affiliation{Laboratori Nazionali di Frascati, Istituto Nazionale di Fisica Nucleare, I-00044 Frascati, Italy}
\author{J.~Antos}
\affiliation{Comenius University, 842 48 Bratislava, Slovakia; Institute of Experimental Physics, 040 01 Kosice, Slovakia}
\author{G.~Apollinari}
\affiliation{Fermi National Accelerator Laboratory, Batavia, Illinois 60510, USA}
\author{J.A.~Appel}
\affiliation{Fermi National Accelerator Laboratory, Batavia, Illinois 60510, USA}
\author{T.~Arisawa}
\affiliation{Waseda University, Tokyo 169, Japan}
\author{A.~Artikov}
\affiliation{Joint Institute for Nuclear Research, RU-141980 Dubna, Russia}
% ...

\collaboration{CDF Collaboration}
\altaffiliation[With visitors from ]{$^a$University of British Columbia, Vancouver, BC V6T 1Z1, Canada,
$^b$Istituto Nazionale di Fisica Nucleare, Sezione di Cagliari, 09042 Monserrato (Cagliari), Italy,
$^c$University of California Irvine, Irvine, CA 92697, USA,
$^e$Institute of Physics, Academy of Sciences of the Czech Republic, 182~21, Czech Republic,
$^f$CERN, CH-1211 Geneva, Switzerland,
$^g$Cornell University, Ithaca, NY 14853, USA,
$^{dd}$The University of Jordan, Amman 11942, Jordan,
$^h$University of Cyprus, Nicosia CY-1678, Cyprus,
$^i$Office of Science, U.S. Department of Energy, Washington, DC 20585, USA,
$^j$University College Dublin, Dublin 4, Ireland,
$^k$ETH, 8092 Z\"{u}rich, Switzerland,
$^l$University of Fukui, Fukui City, Fukui Prefecture, Japan 910-0017,
$^m$Universidad Iberoamericana, Lomas de Santa Fe, M\'{e}xico, C.P. 01219, Distrito Federal,
$^n$University of Iowa, Iowa City, IA 52242, USA,
$^o$Kinki University, Higashi-Osaka City, Japan 577-8502,
$^p$Kansas State University, Manhattan, KS 66506, USA,
$^q$Brookhaven National Laboratory, Upton, NY 11973, USA,
$^r$University of Manchester, Manchester M13 9PL, United Kingdom,
$^s$Queen Mary, University of London, London, E1 4NS, United Kingdom,
$^t$University of Melbourne, Victoria 3010, Australia,
$^u$Muons, Inc., Batavia, IL 60510, USA,
$^v$Nagasaki Institute of Applied Science, Nagasaki 851-0193, Japan,
$^w$National Research Nuclear University, Moscow 115409, Russia,
$^x$Northwestern University, Evanston, IL 60208, USA,
$^y$University of Notre Dame, Notre Dame, IN 46556, USA,
$^z$Universidad de Oviedo, E-33007 Oviedo, Spain,
$^{aa}$CNRS-IN2P3, Paris, F-75205 France,
$^{cc}$Universidad Tecnica Federico Santa Maria, 110v Valparaiso, Chile,
$^{ll}$Universite catholique de Louvain, 1348 Louvain-La-Neuve, Belgium,
$^{oo}$University of Z\"{u}rich, 8006 Z\"{u}rich, Switzerland,
$^{pp}$Massachusetts General Hospital and Harvard Medical School, Boston, MA 02114 USA,
$^{qq}$Hampton University, Hampton, VA 23668, USA,
$^{rr}$Los Alamos National Laboratory, Los Alamos, NM 87544, USA
}
\noaffiliation

%% file: metj_fulldata_v5.3.bbl
\begin{thebibliography}{99}
\bibitem{cordSystem} CDF uses a cylindrical coordinate system with the $z$ axis along the proton beam axis. Pseudorapidity is defined as $\eta \equiv -\ln(\tan(\theta/2))$, where $\theta$ is the polar angle relative to the proton beam direction, and $\phi$ is the azimuthal angle.
\end{thebibliography}
